\documentclass[12pt]{article}
\usepackage{feynmp,epsf}
\def\lnspr{\ln\frac{\displaystyle 1+\beta'}{\displaystyle 1-\beta'}}

\def\lnsbm{\ln\frac{\displaystyle xs}{\displaystyle m^2_\tau}}
\def\lnsm{\ln\frac{\displaystyle s}{\displaystyle m^2_\tau}}
\def\lnsc{\ln\frac{\displaystyle 1}{\displaystyle 1-x}}
\def\ba{\begin{eqnarray}}
\def\ea{\end{eqnarray}}
\begin{document}
\begin{fmffile}{ttgamma}\unitlength=1mm
\thispagestyle{empty}
\onecolumn
\begin{flushleft} 
{\tt
DESY 96-212
\\
hep-ph/9610437
\\
October 1996}
\end{flushleft}

\vfill

\begin{center}{\huge %\bf
 Predictions for Anomalous $\tau^+\tau^-\gamma$ 
\vspace*{.5cm}
\\
Production at LEP~1
\vspace*{3.4cm}
     } % endoftitle

\vfill

\noindent
{\Large Jochen~Biebel and Tord~Riemann}\\
\end{center}
\vspace*{1.0cm}
\begin{center}
\large
%\begin{normalsize}
%   ---1---
%\begin{tabbing}
Deutsches~Elektronensynchrotron~DESY --~Institut~f\"ur~Hochenergiephysik
\\  Platanenallee 6, D-15738 Zeuthen, Germany
%\end{tabbing}
%   ---2---
\end{center}

\vfill

\centerline{\large Abstract}

\bigskip

\normalsize
\noindent
We calculate distributions for $\tau^+\tau^-\gamma$ production at
LEP~1 taking into
account a potentially existing anomalous magnetic moment $a_{\tau}$ of the
$\tau$ lepton.  
The existing upper limits for $|a_{\tau}|$ are known from the
dependence of the decay $Z^0 \to \tau^+\tau^-\gamma$
on $a_{\tau}^2$ and are of the order of ($1 - 5$)\%.
We show that such limits are also sensitive to linear terms in
$a_{\tau}$, which are of equal importance at $|a_{\tau}| \sim (1 - 2)$\% 
 and dominate below this value.
Contributions from an electric dipole moment $d_{\tau}$ do not
interfere with the electromagnetic vertex or with the anomalous
magnetic moment. 
Appropriate formulae are derived.

\vfill

%==============================================================
\newpage
\begin{section}{Introduction}\label{intro}
%----------------------------------------------
There are few basic physical parameters which may be measured best at LEP~1. 
One of them is the anomalous magnetic dipole moment of the $\tau$ lepton.

\bigskip

The matrix element of the electromagnetic current of a particle
with mass $m$ and spin~$\frac{1}{2}$ has the form~\cite{weinberg}:
%----------------------------------
\begin{equation}
e\bar{u}(p')\Biggl[\gamma^\mu
F_1(q^2)+i\frac{F_2(q^2)}{2m}\sigma^{\mu\nu}q_\nu
\Biggr]u(p),
\label{ecurrent} 
\end{equation}
%----------------------------------
with $q=p'-p$ and $\sigma^{\mu\nu} = \frac{i}{2} \, [\gamma^\mu ,
\gamma^\nu ]$.  
The corresponding magnetic dipole moment is given by:
%----------------------------------
\begin{equation}
\mu=\frac{e\hbar}{2m}\left[ F_1(0)+F_2(0)\right] ,
\end{equation}
%----------------------------------
with $F_1(0)=1$. 
The anomalous magnetic dipole moment of a fermion $f$ is defined as
%----------------------------------
\begin{equation}
a_f = F^f_2(0)=   \frac{2m_f}{e\hbar}
\, \mu_f\, - \,1.
 \end{equation}
%----------------------------------
Without radiative corrections or anomalous contributions, it is
$a_f=0$~\cite{dirac}.
The lowest-order QED correction is $a_f = \alpha/(2\pi) =
0.001161$~\cite{schwinger}.   
The magnetic moments of electron and muon are measured with high
precision with the
spin-precession method~\cite{pdg96}:
%------------------------
\begin{eqnarray}
\mu_e&=&(1.001159652193\pm0.000000000010)\frac{e\hbar}{2m_e},
\\
\mu_\mu&=&(1.001165923\pm0.000000008)\frac{e\hbar}{2m_\mu}.
\end{eqnarray}
%----------------------
They are in complete agreement with higher-order theoretical predictions 
and represent an important test of the Standard Model~\cite{jegerlehner}. 

For the short-lived $\tau$ lepton 
the spin-precession method cannot be used.
To our knowledge, the first measurement of the form factor $F_2(q^2)$
of the $\tau$ lepton was performed with the $e^+e^-$ annihilation
experiments at PETRA~\cite{silverman83}.
The result was $F_2(q^2\neq0)\leq0.02$ (95\% CL) for values of $q^2$ in the
range (5 $ - $ 37 GeV)$^2$.
The best limit for $F_2(q^2)$ is due to an analysis of Escribano and
Mass\'o~\cite{escribano93}. 
Using an effective lagrangian approach to the $Z^0$ width, they 
derived from experimental values for the ${\bar \tau}\tau Z^0$ couplings
the limit 
%$F_2(M_Z^2) < 0.01$ (95\% CL).  
$F_2(M_Z^2) < 0.005$ (95\% CL).  
This is one order of magnitude larger than the theoretical
prediction~\cite{samuel}:
%--------------------------
\begin{equation}
a_\tau^{\mbox{\scriptsize th}}=0.001177
\end{equation}
%----------------------
However, since the $q^2$-dependence of $F_2$ is unknown, 
it is impossible to infer the static magnetic moment $a_{\tau}$
from the above measurements of $F_2(q^2)$ at $q^2\neq 0$. 

\bigskip

The first {\em direct} determination of the anomalous magnetic moment of the
$\tau$ lepton, i.e. of $F_2(0)$ 
is due to Grifols and M\'endez~\cite{grifols91} (95\% CL): 
%--
\ba
|a_{\tau}|  \leq 0.11
\label{grilim}
\ea
%---
They derived this limit from a calculation of the quantitative effect
of the electromagnetic current~(\ref{ecurrent}) for on-mass-shell
photons (with $q^2=0$) on the partial $Z^0$ decay width:
%------------------
\begin{eqnarray}
\Gamma\left[Z^0\rightarrow\tau^+\tau^- \gamma \right]
&=& \Gamma_0+
\Gamma_{\mbox{\scriptsize ano}},
\nonumber\\
\Gamma_{\mbox{\scriptsize ano}}
&=& \frac{\alpha^2 a_{\tau}^2
M_Z^3}{1024\pi m^2_\tau\sin^2\theta_W\cos^2\theta_W}\Biggl[(v^2+a^2)-
\frac{1}{9}(v^2-a^2)\Biggr].
\label{grif}
\end{eqnarray}
%----------------------

\begin{figure}[bthp]
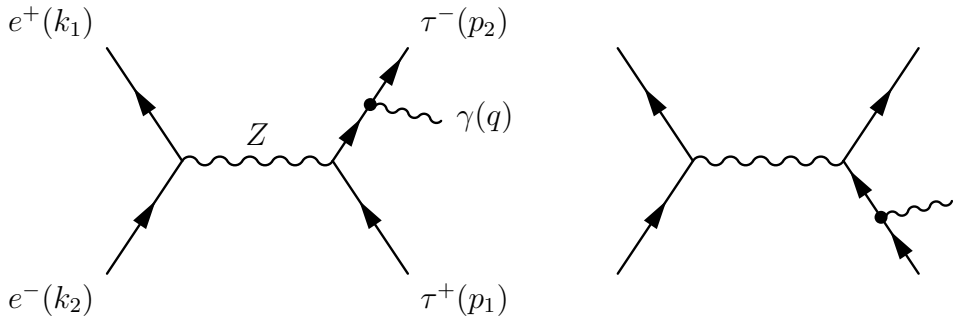

\begin{center}
  \begin{fmfchar*}(50,30)
    \fmfleft{k1,k2}
    \fmfright{p2,k,q,p1}
    \fmf{fermion,label=$$,l.s=left}{k1,v1}
    \fmf{fermion,label=$$,l.s=left}{v1,k2}
    \fmf{fermion,label=$$,l.s=left}{p2,v2}
    \fmf{phantom,label=$$,l.s=left}{v2,p1}
    \fmf{photon,label=$Z$,tension=1,l.s=left}{v1,v2}
	\fmffreeze
    \fmf{fermion}{v2,v3,p1}
	\fmffreeze
    \fmfdot{v3}
    \fmf{photon,label=$$,tension=1,l.s=left}{v3,q}
	\fmflabel{$e^+(k_1)$}{k2}
	\fmflabel{$e^-(k_2)$}{k1}
	\fmflabel{$\tau^+(p_1)$}{p2}
	\fmflabel{$\tau^-(p_2)$}{p1}
	\fmflabel{$\gamma(q)$}{q}
  \end{fmfchar*}
\qquad\qquad
  \begin{fmfchar*}(50,30)
    \fmfleft{k1,k2}
    \fmfright{p2,k,q,p1}
    \fmf{fermion,label=$$,l.s=left}{k1,v1}
    \fmf{fermion,label=$$,l.s=left}{v1,k2}
    \fmf{fermion,label=$$,l.s=left}{v2,p1}
    \fmf{phantom,label=$$,l.s=left}{p2,v2}
    \fmf{photon,label=$$,tension=1,l.s=left}{v1,v2}
	\fmffreeze
    \fmf{fermion}{p2,v3,v2}
	\fmffreeze
    \fmfdot{v3}
    \fmf{photon,label=$$,tension=1,l.s=left}{v3,k}
  \end{fmfchar*}
\end{center}
\caption{\it Diagrams contributing to final-state
radiation with anomalous couplings.\label{diagram1}}
\end{figure}

\noindent
The standard model contribution is $\Gamma_0$ and 
$v=1-4\sin^2\theta_W^{eff}$, $a=1$.

The limit~(\ref{grilim}) has been improved recently by a dedicated
analysis of the L3
collaboration~\cite{l3warsaw,lustermann} (95\% CL):
%--
\ba
|a_{\tau}|  \leq 0.049
\label{lthreelim}
\ea
%---
L3 analyzed the distribution
%------------------
\begin{equation}
\frac{
\mbox{d}\Gamma_{\mbox{\scriptsize ano}}}{\mbox{d}E_\gamma}=
\frac{\alpha^2
a_{\tau}^2
E_\gamma}{144\pi M_Zm^2_\tau\sin^2\theta_W
\cos^2\theta_W}\left[ 3(v^2+a^2)M_Z(M_Z-2E_\gamma)+2E_\gamma^2a^2
\right] .
\label{lthreedist}
\end{equation}
%--------------

As may be seen from the above expressions, both determinations of
$a_{\tau}$ neglect terms linear in $a_{\tau}$.
A simple estimate based on the square of 
matrix elements indicates that the linear terms are
in fact suppressed in 
comparison to terms with $a_{\tau}^2$ 
by a factor $r(s) = m_\tau^2/s$, with $s$ being the scale of the process.
At LEP~1, $s=M_Z^2$ and $r(M_Z^2) \approx 1/2633 = 0.38$ per mil.
Thus, as long as a limit on $|a_{\tau}|$ considerably exceeds this
value, the neglect of linear terms in $a_{\tau}$ seems to be justified.

\bigskip

Recently, a calculation of distributions for the process
%---
\ba
e^+ e^- \rightarrow (Z^0, \gamma) \rightarrow \tau^+ \tau^- \gamma
\label{ttgam}
\ea
%----
at LEP~1 energies gave hints on a stronger influence of the linear
terms in $a_{\tau}^2$ than the above estimate based on $r(M_Z^2)$
suggests~\cite{swaintalk}. 
Because the argumentation is based on a numerical calculation, one may ask
for an analytical re-calculation in order to better
understand whether the effect really exists and what its origin is.
Last but not least one needs an estimate for the range of values of
$a_{\tau}$ where an improved calculation is necessary.  
In addition, besides the total rate and the energy spectrum of
decaying $Z^0$ bosons, the angular distribution of photons and $\tau$ leptons
in the 
center-of-mass system (cms) may be well measured at LEP.
Thus, one should clarify whether these distributions may be of use for
a determination of limits for $a_{\tau}$~\cite{swaintalk}.

\bigskip

%%%{\tt  make a ps file of the feynman diagram} \\
In this article, we calculate several distributions and the total rate
for the reaction~(\ref{ttgam}) arising from the Feynman diagrams of
Figure~\ref{diagram1} with account of the anomalous part of
vertex~(\ref{ecurrent}). 
We formulate the problem exact in $m_{\tau}$, but will later neglect higher
powers of $m_{\tau}$ where they are evidently superfluously for 
numerical estimates.
Further, we can safely neglect anomalous contributions arising from
the initial-final state interferences, from the $\gamma Z^0$
interference, and from $\gamma$ exchange. 
Other Standard Model contributions like e.g. initial-state radiation
are assumed to be known.

In Section~\ref{calc}, we discuss the analytical formulae for
$\mbox{d}^2\sigma / \mbox{d} \cos \theta_{\gamma} \mbox{d}E_{\gamma}$, 
%$\mbox{d}\sigma / \mbox{d} \cos \theta_{\gamma}$,
$\mbox{d} \sigma / \mbox{d}E_{\gamma}$,
$\sigma_{tot}$ .
Modifications arising from an electric dipole moment of the $\tau$
lepton are mentioned.
Section~\ref{discuss} contains a numerical discussion.
The application of simple cuts is also considered.
In Appendix~A the phase space parameterization and the choice of
four-vectors are explained.

\end{section}

%\newpage
%===================================================================
\begin{section}{Distributions for $e^+e^-\rightarrow Z^0\rightarrow
	        \tau^+\tau^-\gamma$}
\label{calc}

%      \input{feyndia}
%-----------------------------------------------------
%this has to be placed after input feyndia
%=====================================================
%The matrix elements of the process $e^+e^-\rightarrow
%Z^0\rightarrow\tau^+\tau^-\gamma$ can be derived with the usual feynman
%rules out of the diagrams shown in figure \ref{diagram1}. 

We used {\tt form}~\cite{form} for the calculation of the square of 
matrix elements, then inserted the products of four-momenta in the
center-of-mass system as determined in Appendix~\ref{psi}, and finally
performed the various integrations. 
The twofold cross-section differential in energy and production angle of the
photon in the cms is:
%-----------------------------
\begin{eqnarray}
   \frac{\mbox{d}^2\sigma}{\mbox{d}x'\mbox{d}
   \cos\theta_\gamma}&=&
   \beta'C\Biggl\{\frac{3}{8}(1-x')\Biggl[(v^2+a^2)
   \frac{1}{\beta'}\lnspr(1+\cos^2\theta_\gamma)-4a^2\Biggr]a_{\tau}
\nonumber \\
\nonumber\\
&&\mbox{}
+ 
\frac{1}{32}(1-x')\Biggl[a^2[(1-x')^2
(1-\cos^2\theta_\gamma)+8x']\vphantom{\lnspr}
\nonumber \\
\nonumber\\
&&\mbox{}\vphantom{\lnspr}+3(v^2-a^2)x'(1+\cos^2
\theta_\gamma)\Biggr] \frac{s}{m_\tau^2} a_{\tau}^2
\nonumber \\   \nonumber\\
&&\mbox{}+\frac{3(v^2+a^2)}{8(1-x')}\Biggl[(1+\cos^2\theta_\gamma)
   (1+x'^2)\frac{1}{\beta'}\lnspr
\nonumber \\
\nonumber\\
&&\mbox{}\vphantom{\lnspr}-(1+\cos^2\theta_\gamma)(1-x')^2
   -8x'\cos^2\theta_\gamma\Biggr]\Biggr\},
\label{2fold}
\end{eqnarray}
%----------------------------
\ba
%C=\frac{4\alpha^3}{3s}(v^2+a^2)\Biggl[\frac{G_\mu M_Z^2} 
%{8\sqrt{2}\pi\alpha}\Biggr]^2\frac{s^2}{\left(s-M^2_Z\right)^2+M^2_Z
%\Gamma^2_Z(s)}.
C
&=&
\frac{\alpha}{\pi}
\frac{G_{\mu}^2 M_Z^4} {96\pi}
\frac{s}{\left|s-M^2_Z+iM_Z\Gamma_Z(s)\right|^2}
(v^2+a^2),
\label{c}
\\ 
\nonumber \\ 
\beta'&=& \sqrt{1-\frac{4m^2_\tau}{x's}},
\ea
%--------------------
and $x'$ is the normalized invariant mass $s'$ of the $\tau^+\tau^-$:
\ba
x' &=& \frac{s'}{s}.
\ea
It is related to the center-of-mass photon energy $E_\gamma$:
%-------------
\begin{equation}
E_\gamma=\frac{\sqrt{s}}{2}(1-x').
\end{equation}
%-------------------
In (\ref{2fold}), some terms proportional to $a_{\tau}$
are enlarged compared to $a_{\tau}^2$ due to a logarithmic factor
$\lnspr \approx \ln(x's/m_{\tau}^2)$, which 
is $7.88$ for $x' \rightarrow 1$ at $\sqrt{s} = M_Z = 91.187$ GeV and
$m_{\tau} = 1.777$ GeV.
The simple numerical factors also disfavor $a_{\tau}^2$ compared to $a_{\tau}$.
Thus, the suppression factor $r(M_Z^2)$ overestimates by far the relation 
between the quadratic and linear terms. 
We should also remark here that the logarithmic term $a_{\tau} \cdot
\ln(s/m_{\tau}^2)$ does not represent a mass singularity.
It arises from the combination $(\sqrt{s} \cdot a_{\tau}/m_{\tau}) \cdot
(m_{\tau}/\sqrt{s}) \ln(s/m_{\tau}^2)$, where  $a_{\tau}/m_{\tau}$ has its
origin in~(\ref{ecurrent}) and the factor
$(m_{\tau}/\sqrt{s})
\ln(s/m_{\tau}^2))$ arises from photonic corrections and vanishes for
small $m_{\tau}/\sqrt{s}$ as it should do. 

As one would expect, distribution~(\ref{2fold}) is smooth in 
$\cos^2\theta_\gamma$ and independent of $\cos\theta_\gamma$.
The Born angular distribution of the $\tau$ lepton is smooth and the
final state photons are preferentially collinear.
The forward-backward asymmetry (in  $\cos\theta_\gamma$) must vanish and thus
odd powers in 
$\cos\theta_\gamma$ are forbidden.

The integration of~(\ref{2fold}) over $\cos\theta_\gamma$ is simple
and yields
%---------------------
\begin{eqnarray}
\frac{\mbox{d}\sigma}{\mbox{d}x'}
&=& 
\beta'  C\Biggl\{\vphantom{\lnspr}
(1-x')\Biggl[(v^2+a^2)
\frac{1}{\beta'}
\lnspr-3a^2\Biggr]a_{\tau}
\nonumber\\  
\nonumber\\  
&&\mbox{}+\frac{1}{4}(1-x')
\Biggl[a^2\frac{(1-x')^2}{6}+(v^2+a^2)x'
   \Biggr] \frac{s}{m_\tau^2} a_{\tau}^2
\nonumber
\\
\nonumber\\  
&&\mbox{}+\frac{v^2+a^2}{1-x'}(1+x'^2)\Biggl(\frac{1}{\beta'}\lnspr-1\Biggr)
   \Biggr\} .
\label{dsxp}
\end{eqnarray}
%-------------------------

The terms proportional to $a_{\tau}^2$ may be found
in~(\ref{lthreedist}) already.  

An integration of~(\ref{dsxp}) with lower cut $x$ on the photon energy,
\ba
\sigma_{hard}(x) &=&
\int_{o}^{x} dx' \frac{\mbox{d}\sigma}{\mbox{d}x'} ,
\ea
with
%-----------------------------
\begin{equation}
x=1-2\frac{E_{\gamma}^{\min}}{\sqrt{s}},
\end{equation}
%-----------------------
yields
%-------------------------------
\begin{eqnarray}
\sigma_{hard}(x)&=&
C\Biggl\{-\frac{x}{2}
\Biggl[(v^2+a^2)\Biggl((x-2)\lnsbm
   +2-\frac{x}{2}\Biggr)-3a^2(x-2)\Biggr]a_{\tau}
\nonumber\\
\nonumber\\
&&\mbox{}+\frac{1}{24}\Biggl[a^2
\frac{1-(1-x)^4}{4}+(v^2+a^2)
  x^2 (3-2x)\Biggr]  \frac{s}{m_\tau^2} a_{\tau}^2
\nonumber\\
\nonumber\\
&&\mbox{}+(v^2+a^2)\Biggl[\frac{3}{4}x^2-\frac{x}{2}
   (2+x)\lnsbm+2x
%-\frac{\pi^2}{3}
   +2 \, \mbox{Li}_2(1-x)
-2 \, \mbox{Li}_2(1)
\nonumber\\
\nonumber\\
&&\mbox{}+2\lnsc\Biggl(\lnsm-1\Biggr)\Biggr]\Biggr\} .
\label{wcut}
\end{eqnarray}
%-----------------------------------
The function $\mbox{Li}_2$ is the Euler dilogarithm,
$\mbox{Li}_2(1)=\pi^2/6.$ 

In all the distributions, it is only the Standard Model part of the
cross-section which is infrared singular for $x\rightarrow1$. 
The anomalous part of the electromagnetic vertex
is proportional to the photon momentum and therefore infrared safe.

To remove the cut, it is necessary to calculate the
soft photon part of the bremsstrahlung contribution.
Furthermore, the photonic vertex correction to $\tau$ pair production
has to be added in order to cancel
the infrared singularity.
The sum of soft and virtual corrections is known (see,
e.g.~\cite{npb}) to read
%---------------
\begin{equation}
%\mbox{SoftVirt}
\sigma_{s+v}(x)
=C(v^2+a^2)\Biggl\{
2\Biggl(\lnsm-1\Biggr)
\Biggl[\ln(1-x) + \frac{3}{4}\Biggr] +2 \, \mbox{Li}_2(1)-\frac{1}{2} 
\Biggr\} .
\label{corrections}
\end{equation}
%---------------------------------

In sum, the ${\cal O}(\alpha^3)$ part of the total cross-section for
$\tau^+ \tau^- (\gamma)$ production is given by
%---------------------------
\begin{eqnarray}
\sigma_{tot}&=&
\sigma_{hard}(x) + \sigma_{s+v}(x)
\nonumber\\
\nonumber\\
&=&
C\Biggl\{(v^2+a^2)\Biggl[ \frac{3}{4}-\frac{1}{2}
	\Biggl(3-\lnsm\Biggr)a_{\tau}
	+\frac{3}{64}\frac{s}{m_\tau^2} a_{\tau}^2 \Biggr]
%\nonumber\\
%\nonumber\\
%&&\mbox{}
+(a^2-v^2)\Biggl(-\frac{3}{4}a_{\tau}
+\frac{1}{192}\frac{s}{m_\tau^2}a_{\tau}^2
	\Biggr)\Biggr\}.
\nonumber\\
\end{eqnarray}
%------------------------
The terms proportional to $a_{\tau}^2$ may be found already in~(\ref{grif}) and
the Standard Model contribution is also well-known.

The existence of a non-vanishing electric dipole moment $d_{\tau}$
would lead to minor modifications of the formulae.
In the matrix element~(\ref{ecurrent}), one has to replace
\ba
i\frac{F_2(q^2)}{2m} &\rightarrow&
i\frac{F_2(q^2)}{2m} + F_3(q^2) \gamma_5,
\\
\nonumber\\
d_{\tau} &=& e \, F_3(0).
\ea
There appear no interferences of $d_{\tau}$ with the electromagnetic
vertex or with $a_{\tau}$ in the cross-sections.
The contributions of the electric dipole moment may be obtained by the
following replacement:
\ba
\frac{a_{\tau}^2}{m_{\tau}^2} \to 
\frac{a_{\tau}^2}{m_{\tau}^2} \, + \, 4 \frac{d_{\tau}^2}{e^2}.
\ea
Since the interferences depend only on $a_{\tau}$, a combined fit to both
$a_{\tau}$ and $d_{\tau}$ becomes possible.
\end{section}
%==================================================================
\begin{section}{Numerical Results}\label{discuss}
%--------------------------------------------------
The numerical calculations have been done with
$\sin^2\theta_W^{eff}=0.2320$.

Figure~\ref{f2} compares the influence of the terms proportional to $a_{\tau}$
and $a_{\tau}^2$.  
An anomalous cross-section contribution of less than about 1.5 pbarn leads to a
limit of about 
$|a_{\tau}|<5\%$ from the $a_{\tau}^2$ terms alone.
The same measurement yields from the complete anomalous contribution a limit of
about $-5.6\% < a_{\tau} < 4.4\%$.
Similarly, a limit of $|a_{\tau}|<3\%$ transforms into about  $-3.6\% <
a_{\tau} < 2.4\%$. 
Evidently, at $|a_{\tau}| \sim 2\%$ and below this value linear terms in
$a_{\tau}$ are indispensible for a correct interpretation of data.

\begin{figure}
  \begin{center}
    \epsfxsize=10cm
    \leavevmode
    \epsffile{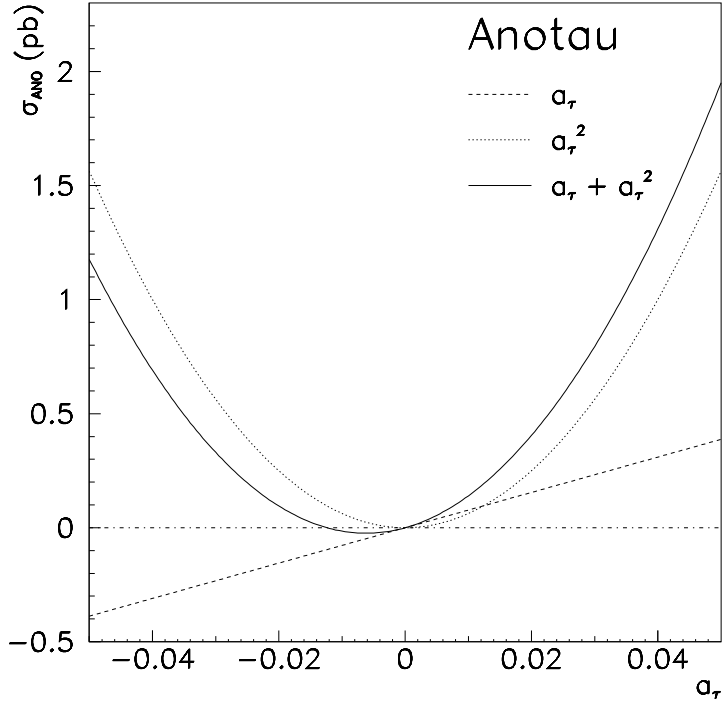}
  \end{center}
  \caption{\it
The contribution of the anomalous terms to $\sigma_{hard}(x)$.
\label{f2}
}
\end{figure}

Figure~\ref{f5} shows the photon energy distribution of the anomalous terms.
The influence of the linear terms is more pronounced than in the integrated
cross-section. 
Due to the $(1-x')$ dependence, the anomalous terms  are proportional to
$E_{\gamma}$ but get damped at high photon energy due to the overall factor
$\beta'$.

Here we have to discuss an important cut of the analysis.
It is impossible to identify photons when emitted parallel to the radiating
$\tau$. 
Thus, a cone has to be cut around the production angles $\theta_{1,2}^*$ 
of the $\tau$ leptons in the cms.
These angles may be related by expressing the scalar products
$p_{1}k$, $p_{2}k$ in two different ways
%-------
\ba
p_{1/2} k &=&
p_{1/2}^0 k^0 - \sum_{i=1}^3 p_{1/2}^i k^i
%\nonumber \\
= p_{1/2}^0 k^0 \left[ 1-\beta_{1/2} \cos \theta_{1/2}^* \right].
\label{cost}
\ea
The momentum components are taken from~(\ref{pot}).
The velocities of $\tau^{+/-}$ in the cms are:
%---
\ba
\beta_{1/2} &=& \sqrt{1-\frac{m_\tau^2}{(p_{1/2}^0)^2}}
=
\left( 1 - \frac{4m_{\tau}^2}{s} 
\frac{1}{
\frac{1}{4}\left[ (1+x') \mp (1-x')\beta'\cos\theta \right]^2
} \right),
\label{betao}
\ea
%--

\noindent
and the angles between $\tau^{+/-}$ and the photon in the cms:
%---  
\ba
\cos\theta_{1/2}^* &=& 
\frac{1}{\beta_{1/2}} \left[ 1- \frac{1 \mp \beta'\cos\theta}
{\frac{1}{2}\left[ (1+x') \mp (1-x')\beta'\cos\theta \right]}
\right].
\label{thetao}
\ea
%----
In the appendix, (\ref{thetao}) is inverted explicitly.

In order to take advantage from this, one has to derive the
double-differential distribution in $s'$ and $\cos\theta$:
%------------
\ba
\frac{d^2 \sigma}{dx' d\cos\theta_1}
&=&
\beta'C\Biggl\{
(1-x')\left[ \frac{1+x'^2}{2(1-x')}
\left(\frac{1}{t_+} +\frac{1}{t_-}\right)
   - \frac{m_{\tau}^2}{s}\left(\frac{1}{t_+^2} +\frac{1}{t_-^2}\right)
-1\right](v^2+a^2)
\nonumber \\
&& + \frac{1-x'}{2}\left[(1-x')  \left(\frac{1}{t_+} +\frac{1}{t_-}\right)
(v^2+a^2) - 6 a^2\right] a_{\tau}
\nonumber \\
&& + \frac{1-x'}{16}
\left[ 4x'(v^2+a^2) + (1-\cos^2\theta_1)(1-x')^2 \right]\frac{s}
{m_{\tau}^2}
a_{\tau}^2 \Biggr\} .
\label{sigcut}
\ea

\noindent
The next integration may be performed with account of the cuts on
$\theta_{1/2}^*$ 
either numerically or, after the above preparations, also analytically. 

The result is shown in figure~\ref{f5}.
The $a_{\tau}^2$ terms are nearly not influenced, while
linear terms are reduced drastically.
This is due to their peaking behavior in forward direction; a behavior which
is even more pronounced for the Standard Model QED contributions.
Nevertheless, the cuts do not change the conclusions of our study.

Finally, figure~\ref{f4} shows the percentage of anomalous cross-section contributions
compared to the Standard Model prediction for $|a_{\tau}|=5\%$.
If the minimal photon energy is larger, the weight of them rises (but the
number of events goes down).
For negative $a_{\tau}$, the rate is only 60\% of that for positive $a_{\tau}$.

\begin{figure}[tbhp]
  \begin{center}
    \epsfxsize=10cm
    \leavevmode
    \epsffile{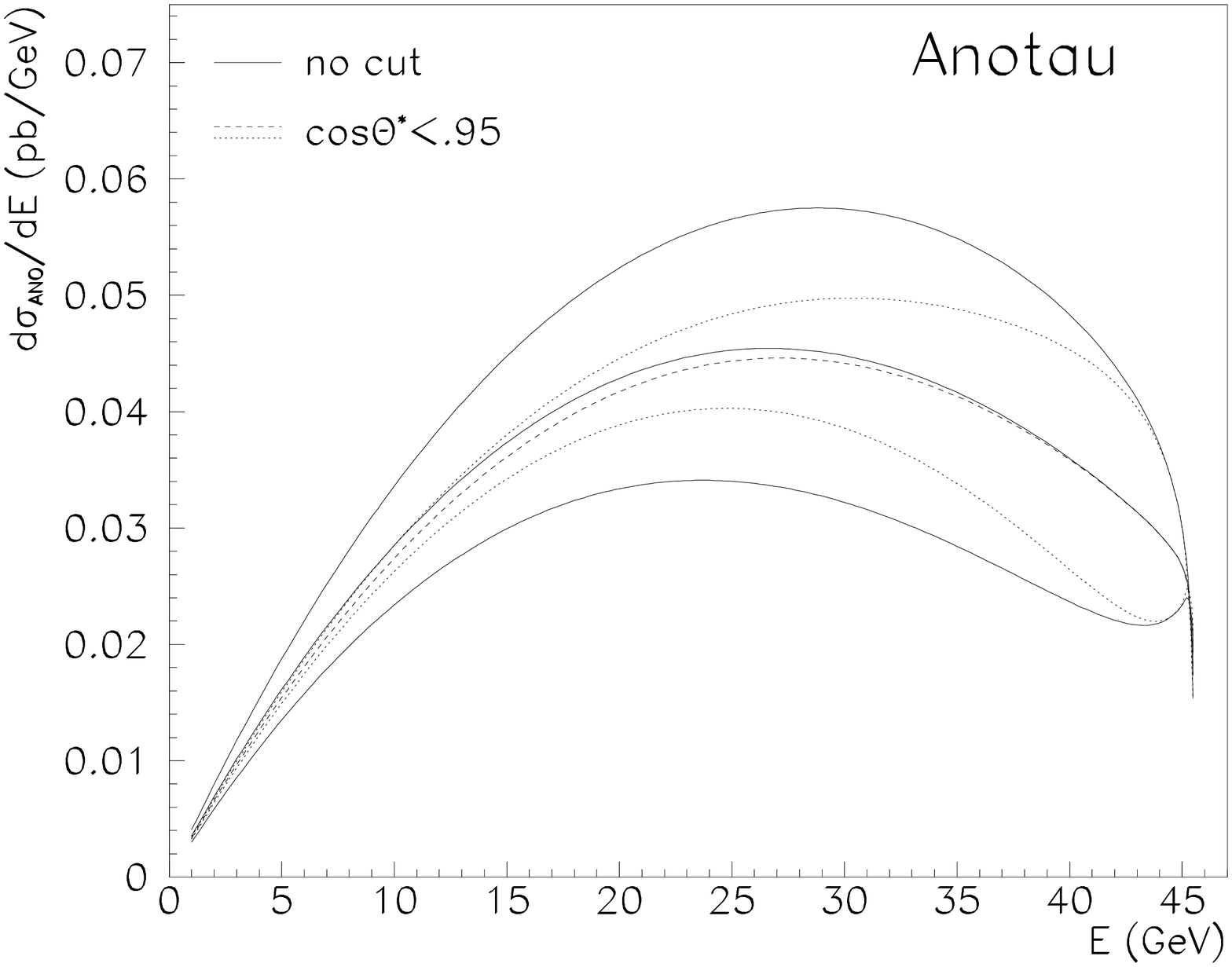}
  \end{center}
  \begin{center}
    \epsfxsize=10cm
    \leavevmode
    \epsffile{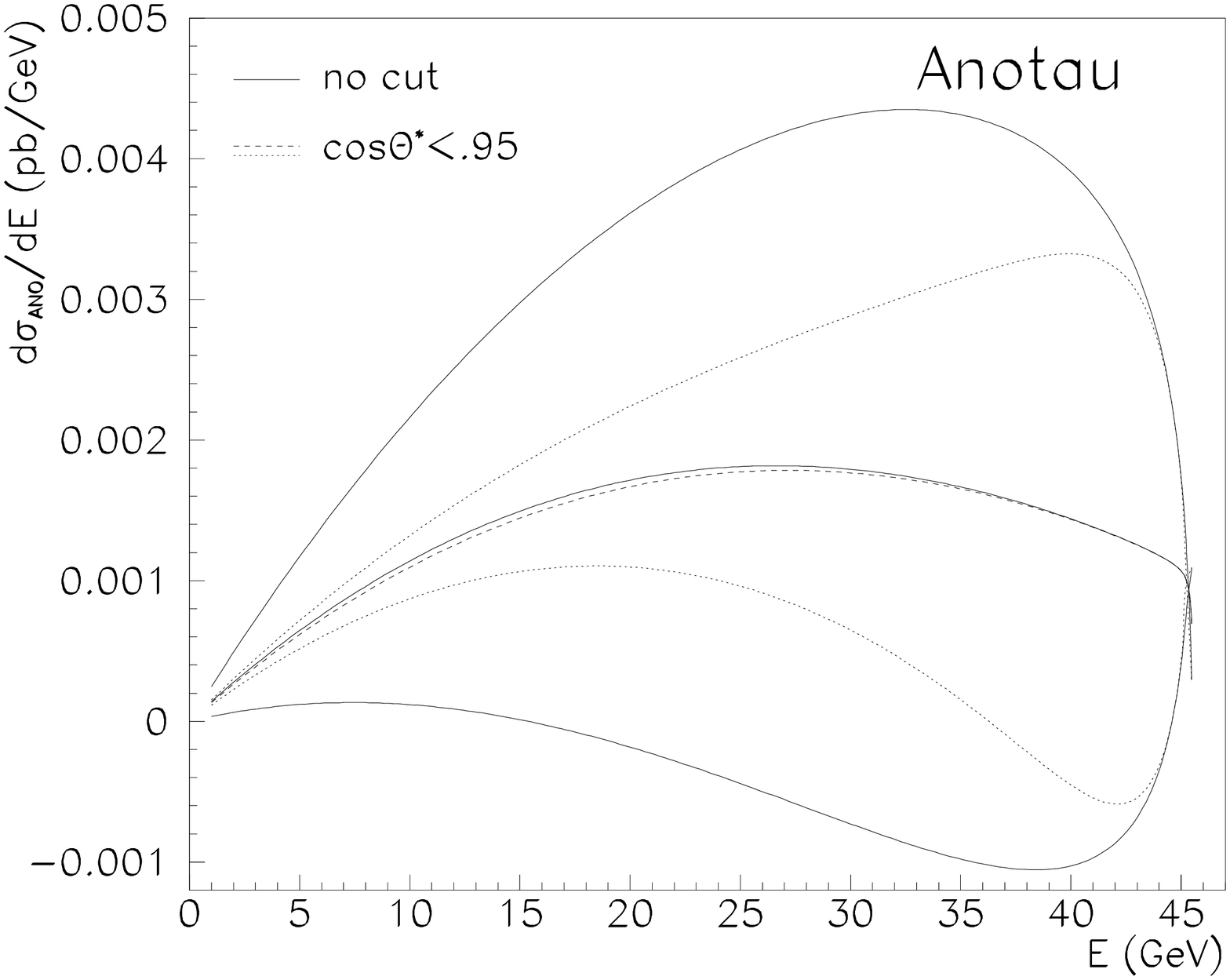}
  \end{center}
  \caption{\it
In the upper figure the two (solid and dashed) lines in the middle
are contributions of the 
anomalous terms quadratic in $a_{\tau}$ for $|a_{\tau}|=5$\%. The two
(solid and dotted) 
lines above are the net anomalous contribution for 
$a_{\tau}=+5$\% and the two (solid and dotted) lines below for
$a_{\tau}=-5$\%. In the lower figure we plot the same curves
for $|a_\tau|=1$\%.      
\label{f5}
}
\end{figure}

\clearpage

%\newpage

%\bigskip

%----
\begin{figure}[tbhp]
  \begin{center}
    \epsfxsize=10cm
    \leavevmode
    \epsffile{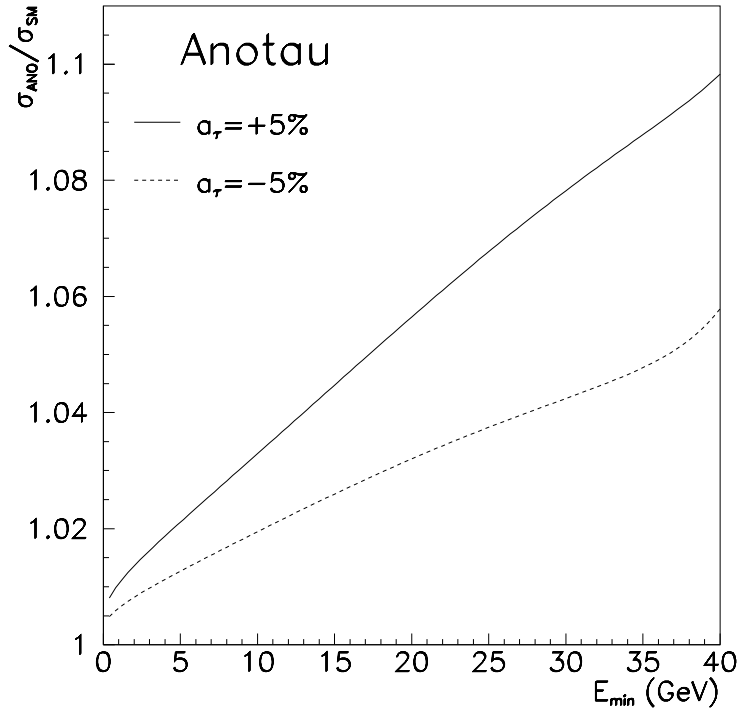}
  \end{center}
  \caption{\it
The relative contribution of anomalous terms compared to the Standard Model
prediction, $\sigma_{hard}^{anom}(x) / \sigma_{hard}^{SM}(x)$, for fixed
$a_{\tau}$, 
as a function of the cut on the minimal photon energy $E_{\min}$. 
\label{f4}    }
\end{figure}

To summarize, 
based on an analytical approach,
we have shown that linear contributions in $a_{\tau}$ are of
numerical relevance 
if the limits on the anomalous magnetic moment of the $\tau$ are
better than a few per cent. Thus,
they have to be taken into account with the present LEP~1 data.
A combined fit to both $a_{\tau}$ and $d_{\tau}$ becomes possible.

\end{section}

\section*{Acknowledgments}
We would like to thank W.~Lohmann and the L3 collaboration for drawing
our attention to the 
problem and to F.~Jegerlehner, W.~Lohmann, W.~Lustermann, M.~Pohl and
J.~Swain for discussions. 

\newpage
%==================================================================
\appendix
\def\theequation{\Alph{section}.\arabic{equation}}
\setcounter{equation}{0}
\begin{section}{Phase Space and Integrals}\label{psi}
%==================================================================
%In this appendix we show the phase space and some integrals used
%in the calculation of the given results. 
Here we will present the
parameterizations of phase space and four-momenta.
We take the ultra-relativistic limit in the electron mass while
$m_{\tau}$ is taken into account exactly if not stated otherwise.

The phase space depends on five independent variables
besides $s$:
The spherical angle of the photon in the cms,
the spherical angle of the $\tau^+$ in the rest system of the
$\tau$-pair, and $x'=s'/s$, $s'$ being the invariant mass of the
$\tau$-pair. 

With these variables, the phase space for the three particle final
state may be expressed  as a
sequence of two two-particle phase spaces:
%------------------------------
\begin{eqnarray}
\mbox{d}\Gamma&=&\frac{\mbox{d}^3p_1}{(2\pi)^32p_1^0}
\frac{\mbox{d}^3p_2}{(2\pi)^32p_2^0}\frac{\mbox{d}^3k}{(2\pi)^32k^0}
\delta^4(k_1+k_2-p_1-p_2-k)
\nonumber
\\
&=&\frac{s}{(2\pi)^9} \, \mbox{d}\Gamma_{\tau} \, \mbox{d}
\Gamma_c \, \mbox{d}x'\vphantom{\frac{2}{2}},
\label{phase}
\end{eqnarray}
%---
with
%---
\begin{eqnarray}
\mbox{d}\Gamma_{\tau}&=&\frac{\sqrt{\lambda(x',m^2_\tau/s,m^2_\tau/s)}}{8x'}
\mbox{d}\varphi_1\mbox{d}\cos\theta_1=\frac{\beta'}{8}
\mbox{d}\varphi_1\mbox{d}\cos\theta_1,
\\ \nonumber \\ 
\mbox{d}\Gamma_c&=&\frac{\sqrt{\lambda(1,x',0)}}
{8}\mbox{d}\varphi_\gamma\mbox{d}\cos\theta_\gamma,
\end{eqnarray}
%--
and
%---
\begin{eqnarray}
x'&=&\frac{(p_1+p_2)^2}{s},
\\ 
\nonumber \\ 
\beta'&=& \sqrt{1-\frac{4m^2_\tau}{x's}}.
\end{eqnarray}
The $\lambda$ function is defined as usually:
%---------------
\begin{equation}
\lambda(a,b,c)=a^2+b^2+c^2-2ab-2ac-2bc.
\end{equation}
%--------------
%----------------------------

The kinematical limits are:
%--------
\begin{eqnarray}
0 \leq &\varphi_1, \varphi_{\gamma}& \leq 2 \pi,
\\ \nonumber \\
0 \leq &\theta_1, \theta_{\gamma}& \leq  \pi, 
\\ \nonumber \\ 
\frac{4 m_{\tau}^2}{s} \leq & x' & \leq 1.
\label{limits}
\end{eqnarray}
%--------------
The limits on $x'$ are related to those on the photon energy:
%--------
\begin{eqnarray}
0 \leq E_{\gamma} \leq E_{\gamma}^{\max} = \frac{\sqrt{s}}{2} \left( 1
- \frac{4 m_{\tau}^2}{s} \right).
\label{limitg}
\end{eqnarray}
%--------------

A dedicated construction of the four-momenta is crucial for the
subsequent steps of the calculation.
We 
%follow closely the approach which is derived in~\cite{CC11} and
determine all momenta in the cms. 
and %We 
begin with an ansatz for the photon momentum:
%--------------
\begin{equation}
q=\frac{\sqrt{s}}{2}\left(1-x',0,0,1-x'\right).
\end{equation}
%--------------
Then, 
we introduce the photon production angle $\theta_{\gamma}$ as the
angle between the three-momenta of photon and electron in the cms:
%--------------
\begin{eqnarray}
k_1&=&\frac{\sqrt{s}}{2}\left(1,\sin\theta_\gamma,0,-\cos\theta_\gamma\right)
\\ \nonumber \\ 
k_2&=&\frac{\sqrt{s}}{2}\left(1,-\sin\theta_\gamma,0,\cos\theta_\gamma\right).
\end{eqnarray}
%----------------

All the four-momenta are independent of $\varphi_\gamma$.

Finally, we have some freedom to choose the momenta of the
$\tau$ leptons in the  rest system of the $\tau$-pair.  
Following the arguments given in Appendix~B of~\cite{CC11}, we get
after a Lorentz transformation into the cms:
%-------------------- 
\begin{equation}
p_{1/2}=\frac{\sqrt{s}}{4}\Biggl(\mp a(1-x')+1+x',\pm b\cos\varphi_1,
\pm b\sin\varphi_1,\pm a(1+x')-1+x'\Biggr),
\label{pot}
\end{equation}
%----------------------------
\begin{eqnarray}
a&=&\beta'\cos\theta_1,
\\ \nonumber \\ 
b&=&2\sqrt{x'}\beta'\sin\theta_1.
\end{eqnarray}
%----------------------------
The denominators of the $\tau$ propagators are: 
%----------------------------
\begin{eqnarray}
t_+=2p_1q&=&\frac{s}{2}(1-x')(1-\beta'\cos\theta_1),
\\ \nonumber \\ 
t_-=2p_2q&=&\frac{s}{2}(1-x')(1+\beta'\cos\theta_1).
\end{eqnarray}
%----------------------------

The first integrations are over $\varphi_1$ and
$\cos\theta_\gamma$ and are simply polynomial.
The only non-polynomial angular integrations
are over $\theta_1$:
%------------------------
\begin{eqnarray}
\frac{1}{2}\int\limits_{-1}^{1}\mbox{d}\cos\theta_1\frac{m^2_\tau}{t_\pm^2}
&=&
\frac{{x'}}{s(1-{x'})^2},
\label{noc1}
\\ \nonumber \\ 
\frac{1}{2}\int\limits_{-1}^{1}\mbox{d}\cos\theta_1\frac{1}{t_\pm}
&=&
\frac{1}{\beta's(1-x')}\ln\frac{1+\beta'}{1-\beta'}
%%%%\nonumber \\ \nonumber\\
%%%%&\approx&\frac{1}{\beta' s(1-{x'})}\lnspr .
\approx\frac{1}{\beta' s(1-{x'})}\ln\frac{x's}{m^2_\tau} .
\label{noc2}
\end{eqnarray}
%-----------------------
The approximation in~(\ref{noc2}) is used only for the integration
over $x'$ where it does not lead to substantial inaccuracies.
For the integrations with angular cut we have 
to invert (\ref{thetao}).
Equation~(\ref{pot}) may be expressed explicitly in terms 
of $\cos\theta^*_{1/2}$:  
%-----------------------
\begin{equation}
p_{1/2}=|\vec{p}_{1/2}|\left(\frac{1}{\beta_{1/2}},
\sin\theta^*_{1/2}\cos\varphi^*_{1/2},\sin\theta^*_{1/2}\sin\varphi^*_{1/2},
\cos\theta^*_{1/2}\right) ,
\label{pot2}
\end{equation}
%-----------------------
with
\begin{equation}
\cos\theta^*_{1/2}=\frac{p_{1/2}^z}{|\vec{p}_{1/2}|} .
\end{equation}
{From} $p_{1/2}^z$ and 
$|\vec{p}_{1/2}|=\sqrt{{p^x}^2_{1/2}+{p^y}^2_{1/2}+{p^z}^2_{1/2}}$
from (\ref{pot}), we find immediately 
%----
\begin{equation}
\cos\theta^*_{1/2}=\frac{\pm A\cos\theta_1-B}{\sqrt{D(1-\cos^2\theta_1)
             +A^2\cos^2\theta_1\mp2AB\cos\theta_1+B^2}},
\label{thetastar}
\end{equation}
with
\begin{eqnarray}
A&=&\beta'(1+x'),
\\ \nonumber \\ 
B&=&1-x',
\\ \nonumber \\ 
D&=&4x'\beta'^2,
\\ \nonumber \\ 
E&=&A^2-(A^2-D)\cos^2\theta^*_{1/2}
\end{eqnarray}
The inversion of~(\ref{thetastar}) leads to a quadratic equation for
$\cos\theta_{1/2}$ with two solutions:
\begin{equation}
\cos\theta_1=\frac{AB(1-\cos^2\theta_1^*)\pm\sqrt{A^2B^2(1-\cos^2\theta^*_1
)^2-E(B^2-(B^2+D)\cos^2\theta_1^*)}}{E}
\label{theta}
\end{equation}
In the limit $x'\rightarrow1$ (no boost) the
corresponding angles in both systems must be equal. 
Therefore, the
sign in front of the square root has to agree with the sign
of $\cos\theta^*_{1/2}$.

The cuts on $\cos\theta^*$ lead to a reduction of the range of 
integration over $\cos\theta_1$. 
With (\ref{theta}) the new integration
limits are known as functions of $\cos\theta^*$. 
Since one has to apply cuts on both $\tau$ leptons, the integration limits
remain symmetric and the following modifications
of~(\ref{noc1})--(\ref{noc2}) arise: 
%------------------------
\begin{eqnarray}
\frac{1}{2}\int\limits_{-c}^{c}\mbox{d}\cos\theta_1\frac{m^2_{\tau}}{t_\pm^2}
&=& {\cal O}(\frac{m_{\tau}^2}{s})  \rightarrow 0,
\\ \nonumber \\
\frac{1}{2}\int\limits_{-c}^{c}\mbox{d}\cos\theta_1\frac{1}{t_\pm}
&=&
\frac{1}{(1-x')s}\ln\frac{1+c}{1-c}.
\end{eqnarray}
%-----------------------

The only potentially non-trivial integrals are those over $x'$.
It is here where the limits $\beta' \to 1$ and $m_{\tau}^2/x' \to 0$
lead to simplifications.  
Aside from integrals over $(x')^n, (x')^n \ln x'$, there are two
integrals which require an infrared cutoff $x$: 
%---------------------------
\begin{eqnarray}
 \int\limits_{0}^{x}\frac{\ln(x's/m_{\tau}^2)} {1-x'}\mbox{d}x'
&=&\lnsm\lnsc
   -\mbox{Li}_2(1)+\mbox{Li}_2(1-x),
\\
   \int\limits_{0}^{x}\frac{1}{1-x'}\mbox{d}x'&=&
   \ln\frac{1}{1-x} .
%
% \\\int\limits_{0}^{x}{s'}^n\lnspr\mbox{d}x'&=&\frac{x^{n+1}}{n+1}\left(
%	\lnsbm-\frac{1}{n+1}\right)\\
%
%   \int\limits_{0}^{x}{x'}^n\mbox{d}x'&=&\frac{x^{n+1}}{n+1}   
\end{eqnarray}

%----------------------
%The integration limit $x < 1$ prevents the inclusion of soft photons
%into the cross-section and retains it infrared finite.
\end{section}
\end{fmffile}

\newpage
%---------------------------------------------------

\end{document}